\documentclass{ws-ijmpcs}
\usepackage{graphicx}% Include figure files
\usepackage{epstopdf}
\usepackage{dcolumn}% Align table columns on decimal point
\usepackage{float}
\usepackage[normalem]{ulem}
\usepackage{color}
\newcommand{\be}{\begin{equation}}
\newcommand{\ee}{\end{equation}}
\newcommand{\bea}{\begin{eqnarray}}
\newcommand{\eea}{\end{eqnarray}}

\newcommand{\bfk}{\mbox{\boldmath $k$}}

\def\trimmarks{}

\begin{document}

\markboth{Anuradha Misra}
{SSA in electroproduction of $J/\psi$ and QCD-evolved TMD's}

\catchline{}{}{}{}{}
%
%%%%%%%%%%%%%%%%%%%%%%%%%%%%%%%%%%%%%%%%%%%%%%%%%%%%%%%%%%%%%%%%%%%%

\title{Single Spin Asymmetry in Electroproduction of $J/\psi$ and QCD-evolved TMD's}

\author{Rohini M. Godbole}

\address{Centre for High Energy Physics\\
{\it Bangalore, India-560012}\\
rohini@cts.iisc.ernet.in}

\author{Abhiram Kaushik}

\address{Centre for High Energy Physics\\
{\it Bangalore, India-560012}\\
abhiramb@cts.iisc.ernet.in}

\author{Anuradha Misra}

\address{Department of Physics, University of Mumbai\\
{\it Mumbai, India-400098}\\
misra@physics.mu.ac.in}

\author{Vaibhav S. Rawoot}

\address{Institute of Mathematical Sciences\\
{\it Chennai, India}\\
vaibhavrawoot@gmail.com}

\maketitle

\begin{history}
\received{Day Month Year}
\revised{Day Month Year}
% \published{Day Month Year}
\end{history}

\begin{abstract}
We estimate Sivers asymmetry in  low virtuality photoproduction of $J/\psi$ using color evaporation model and taking into account 
$Q^2$- evolution of transverse momentum dependent PDF's and Sivers function. There is a substantial reduction in asymmetry as compared to our previous analysis wherein 
the $Q^2$-dependance came only from DGLAP evolution of collinear part of TMDs. The estimates of asymmetry are comparable to our earlier estimates in which 
we had used analytical solution of only an approximated form of the evolution equations. 
 We have also estimated asymmetry  using 
the latest parametrization by Echevarria {\it et al.} which are based on an evolution kernel in which 
the perturbative part is resummed to NLL accuracy. 
\keywords{Charmonium; TMD PDF's; QCD Evolution.}
\end{abstract}

\ccode{PACS numbers:13.88+e, 13.60.-r, 14.40.Lb, 29.25.Pj.}

\section{Introduction}
	
%Heavy quarkonium systems provide a unique lab for studying the interplay between 
%perturbative and non-perturbative regimes of strong interaction. Quarkonium  is a bound state of 
%a heavy quark $Q$ and its anti-quark ${\bar Q}$ which are initially produced in a perturbative process and then 
%evolve into a bound state non-perturbatively. There are 
The issue of quarkonium production mechanism is an open 
question as none of theoretical  models  which are used to describe the
non-perturbative transformation of the $Q{\bar Q}$ pair into quarkonium i.e. the Color Singlet Modeal (CSM)\cite{Einhorn}\cdash\cite{Lansberg2}, 
Color Evaporation Model (CEM)\cite{fri} 
 and Non-Relativistic 
Quantum Chromodynamics (NRQCD)\cite{Bodwin:1994jh},  is able to explain satisfactorily all the data on both production cross section and polarization measurements. 
Thus, independent tests other than polarization measurements are needed to compare the different production mechanisms. 
One such possible test is transverse single spin asymmetry (SSA)  in charmonium production\cite{Godbole:2012bx,brodsky1,brodsky2} since the  asymmetry in heavy quarkonium production is very sensitive to the production mechanism\cite{Yuan:2008vn}.

 One of the theoretical approaches that has been used to explain these asymmetries in Semi-Inclusive Deep Inelastic Scattering (SIDIS) and Drell Yan (DY) processes is based on a transverse momentum dependent factorization scheme\cite{fact,Collins:2011book} which involves transverse momentum dependent 
parton densities and fragmentation functions collectively referred to as TMD's. One such TMD of great interest is Sivers function which 
gives the probability of finding an unpolarized parton inside a transversly polarized nucleon. 

 In an earlier work, we proposed that transverse SSA in charmonium production can be used to study Sivers effect\cite{Godbole:2012bx}. We presented first estimates of SSA in photoproduction (i.e. low virtuality electroproduction) of charmonium in scattering of 
electrons off transversely polarized protons using CEM. 
In the process that we considered, at LO, there is contribution only from a single
partonic subprocess and hence, it can be used as a clean probe of gluon Sivers function.
Subsequently, we improved our estimates taking into account TMD evolution of TMD PDF's and the Sivers function\cite{Godbole:2013bca}. In present work, we present 
further improved estimates calculated using the latest fits to TMD PDF's given by Echevarria {\it et al.}\cite{Echevarria:2014xaa} and compare them with our earlier estimates. 

\section{Transverse Single Spin Asymmetry in $e + p^\uparrow \rightarrow J/\psi +X$}
We will use a generalization of CEM for our estimates of asymmetry in photoproduction (low virtuality electroproduction) 
of $J/\psi$ by 
taking into account the transverse momentum dependence
of the WW function and the  gluon distribution function \cite{Godbole:2012bx}

\begin{eqnarray}
\sigma^{e+p^\uparrow\rightarrow e+J/\psi + X}=\int_{4m_c^2}^{4m_D^2} dM_{c\bar c}^2 dx_\gamma dx_g [d^2{\bf k}_{\perp\gamma}d^2{\bf k}_{\perp g}]
f_{g/p^{\uparrow}}(x_{g},{\bf k}_{\perp g})  \nonumber \\
\times f_{\gamma/e}(x_{\gamma},{\bf k}_{\perp\gamma})
\frac{d\hat{\sigma}^{\gamma g\rightarrow c\bar{c}}}{dM_{c\bar c}^2} 
%\label{dxec-ep}
\end{eqnarray}
where $f_{\gamma/e}(y,E)$ is the William Weizsacker function\cite{Kniehl} which gives distribution function of the photon in the electron.  
We assume $k_\perp$ dependence of pdf's to be 
factorized in gaussian form \cite{Anselmino:2005ea}
\begin{equation}
f(x,k_{\perp})=f(x)\frac{1}{\pi\langle k^{2}_{\perp}\rangle} 
e^{-k^{2}_{\perp}/\langle{k^{2}_{\perp}\rangle}} 
\end{equation}
with $\langle k^{2}_{\perp}\rangle=0.25 GeV^2 $. For the ${\bf k}_\perp$ dependent WW function also, we use a Gaussian form. Expression for the numerator of the asymmetry is\cite{Godbole:2012bx}
\begin{eqnarray}
\frac{d^4 \sigma^\uparrow}{dyd^2{\bf q}_T}-\frac{d^4\sigma^\downarrow}{dy d^2 {\bf q}_T}=
\frac{1}{2}\int_{4m^2_c}^{4m^2_D}[dM^{2}] \int [dx_{\gamma}dx_{g}d^2{\bf k}_{\perp\gamma} d^2{\bf k}_{\perp g}]
\Delta^{N}f_{g/p^{\uparrow}}(x_{g},{\bf k}_{\perp g})  \nonumber \\
\times f_{\gamma/e}(x_{\gamma},{\bf k}_{\perp\gamma})  
\delta^{4}(p_{g}+p_{\gamma}-q)\>
\hat\sigma_{0}^{\gamma g\rightarrow c\bar{c}}(M^2) 
\label{nssa}
\end{eqnarray}
where $q=p_c+p_{\bar c}$ and the ${{\hat \sigma}_{0}}^{\gamma g\rightarrow c\bar{c}}(M^2)$ is the partonic cross section in lowest order (LO). 

In Eq.(3), $ \Delta^Nf_{g/p^\uparrow}(x,{\bf k}_\perp)$ is the gluon Sivers function for which we 
use the following parametrization\cite{Anselmino1}
\begin{equation}
 \Delta^Nf_{g/p^\uparrow}(x,k_\perp) = 2 {\mathcal N}_g(x)\sqrt{2e}\,\frac{{k_\perp}}{M_{1}}\,e^{-{{k_\perp}^2}/{M_{1}^2}} f_{g/p}(x)
\frac{e^{-k^{2}_{\bot}/\langle{k^{2}_{\bot}\rangle}}}{\pi\langle k^{2}_{\perp}\rangle} 
({\bf S}\cdot {\hat {\bf p}} \times {\hat{\bf k}}_\perp) 
\label{dnf}
\end{equation}
Here, ${\mathcal N}_g(x)$ is the x dependent normalization for which we have used ${\mathcal N}_g(x)=\left( {\mathcal N}_u(x)+
{\mathcal N}_d(x) \right)/2 $\cite{Boer-PRD69(2004)094025}.
Since there is not enough data available to parametrize 
the gluon Sivers function, we have expressed gluon Sivers function in terms of quark Sivers function. For quark Sivers functions, 
we use the following normalization\cite{}
  \begin{equation}
{\mathcal N}_f(x) = N_f x^{a_f} (1-x)^{b_f} \frac{(a_f + b_f)^{(a_f +
b_f)}}{{a_f}^{a_f} {b_f}^{b_f}} \nonumber
\label{siversx} 
\end{equation}
where $a_f, b_f$ and $N_f$ are best fit parameters. %and in Model II
%\begin{equation}
%h(k_\perp)=\frac{2k_\perp M_0}{{k_\perp}^2+M_0^{2}} 
%\end{equation}
%where $M_0=\sqrt{\langle{k_\perp^2}\rangle}$ and $M_1$ are best fit parameters.

%\be
%\bfq_T = q_T(\cos\phi_{q}, \, \sin\phi_{q},\, 0) \quad\quad\quad
%\bfk_{\perp} = k_{\perp}(\cos\phi_{k_\perp}, \, \sin\phi_{k_\perp}, \, 0) \nonumber
%\ee
%The mixed product ${\bfS} \cdot (\hat{\bfp} \times \hat{\bfk}_{\perp})$
%leading to an an azimuthal dependence
%${\bf S}\cdot (\hat{\bf p}\times \hat{\bf k}_\perp)=\cos{\phi_{k_\perp}} $, 
%where $\phi_{k_\perp}$ is the angle that transverse momentum of the parton
%$k_\perp$ makes with x axis.
Taking $sin(\phi_{q}-\phi_S)$  as a weight, the asymmetry integrated over the azimuthal angle of $J/\psi$ is
\begin{equation}
A_N=\frac{\int d\phi_{q}[\int_{4m^2_c}^{4m^2_D}[dM^{2}]\int[d^2{\bf k}_{\perp g}]
\Delta^{N}f_{g/p^\uparrow}(x_{g},{\bf k}_{\perp g})
f_{\gamma/e}(x_{\gamma},{\bf q}_T-{\bf k}_{\perp g})
\hat\sigma_{0}]sin(\phi_{q}-\phi_S)}
{2\int d\phi_{q}[\int_{4m^2_c}^{4m^2_D}[dM^{2}]\int[d^{2}{\bf k}_{\perp g}]
f_{g/P}(x_g,{\bf k}_{\perp g})
f_{\gamma/e}(x_{\gamma},{\bf q}_T-{\bf k}_{\perp g})
\hat\sigma_{0}]}
\label{an2}
\end{equation}
where $\phi_q$ and $\phi_{k_\perp}$ are the azimuthal angles of the transverse momenta  $q_T$ and $k_\perp$
and $x_{g,\gamma} = \frac{M}{\sqrt s} \, e^{\pm y}  $.

\section{QCD  Evolution of TMD PDF's }

Initial phenomenological fits of the Sivers function and other TMD's used TMDs which do not 
evolve with the scale of the process\cite{Collins1,Anselmino1}.
Our initial estimates of Sivers asymmetry were based on these parameters and the TMDs used were 
evolved using DGLAP evolution wherein only the collinear part evolves and the Gaussian width of the transverse 
part is assumed to be fixed. In recent years, the TMD factorization has been derived and implemented 
\cite{Collins:2011book,Aybat:2011zv,Aybat:2011ge}.
% TMD evolution is more complicated as compared to its collinear counterpart 
%because unlike collinear distributions TMDs have rapidity divergences in addition to collinear singularities. 
%TMD evolution describes how the form of the  distribution  and also the 
%width  in momentum space changes with scale. 

A strategy to extract Sivers function from SIDIS data taking into account the TMD $Q^2$ evolution was 
 proposed by Anselmino {\it et al.}\cite{Anselmino:2012aa}. 
 We  have also estimated SSA in electroproduction of $J/\psi$ production based on this strategy
\cite{Godbole:2013bca}. 
%It was found that there is a substantial reduction in asymmetry as compared to our previous analysis wherein 
%the $Q^2$-dependance came only from DGLAP evolution of collinear part of TMDs. However, in Ref. \refcite{Godbole:2013bca}, 
%we have used analytical solution of only an approximated form of the evolution equations. 
In present work, we compare our earlier estimates with improved estimates obtained using exact solution of 
evolution equation. In addition, we have also estimated asymmetry  using 
the latest parametrization by Echevarria {\it et al.}\cite{Echevarria:2014xaa} which are based on an evolution kernel in which 
the perturbative part is resummed to NLL accuracy. 

The energy evolution of a general TMD $F(x,k_\perp,Q)$ is more naturally described in b-space. 
The b-space TMD's   evolves with Q according to 
\begin{equation}
 F(x,b,Q_f)=F(x,b,Q_i)R_{pert}(Q_f,Q_i,b_*)R_{NP}(Q_f,Q_i,b)
\label{evolution}
\end{equation}
where $R_{pert}$ is the  perturbative part of the evolution kernel, 
$R_{NP}$ is the non-perturbative part and $b_*=b/\sqrt{1+(b/b_\text{max})^2}$.
 The perturbative part is given by
\begin{equation}
R(Q_f,Q_i,b)=\exp\left\{-\int_{Q_i}^{Q_f}\frac{d\mu}{\mu}\left(A\ln\frac{Q_f^2}{\mu^2}+B\right)\right\}\left(\frac{Q_f^2}{Q_i^2}\right)^{-D(b;Q_i)}
\end{equation}
where $\frac{dD}{d\ln\mu}=\Gamma_\text{cusp}$
The anomalous dimensions $A$ and $B$ are known up to three loop level\cite{Idilbi:2006dg}. 
The non-perturbative exponential part contains a Q-dependent factor universal to all TMDs and a factor which gives the gaussian width in $b$-space of the particular TMD
\begin{equation}
R_{NP}=\exp\left\{-b^2\left(g_1^\text{TMD}+\frac{g_2}{2}\ln\frac{Q_f}{Q_i}\right)\right\}  
\end{equation}
The $b_*$ prescription  stitches  together the perturbative part(which is valid at low $b$) and non-perturbative part(which is valid at large b.  
$Q^2$-dependent  TMD's  in momentum space are obtained by Fourier transforming $F(x,b,Q_f)$.

\section{Approximate Analytical versus Exact Solution of TMD Evolution Equations}

In the analytical approach of Anselmino {\it et al.} \cite{Anselmino:2012aa}
, one assumes that the kernel $R(Q, Q_0, b)$,   which drives the $Q^2$-evolution of TMD's, becomes independent of b in large b limit, 
i.e. as $b \rightarrow \infty$ , $R(Q, Q_0, b )\rightarrow R(Q, Q_0)$. 
 $b$  integration can then be performed analytically and $Q^2$ dependent PDF's can be obtained. 
 In our earlier work\cite{}, we used $Q^2$-evolved TMD PDF's obtained using this  "approximate, analytical" approach.  We will now compare these results with estimates obtained using exact solution of TMD evolution equations which can be obtained by solving the TMD evolution equation numerically\cite{Anselmino:2012aa}. 

Recently, Echevarria {\it et al.}\cite{Echevarria:2014xaa} have 
considered solution of TMD evolution equations up to NLL accuracy and have performed a global fitting of all 
experimental data on the Sivers asymmetry in SIDIS using this formalism. 
%The perturbative part of the evolution kernel is expressed by Eq.(17) and the non-perturbative part is given by Eq.(18). 
%$Q^2$-evolution of TMD PDF and the derivative of Sivers function in b- space is then given by Eq.(17). 
% \cite{Anselmino:2012aa},
Since the  derivative of b-space  Sivers function satisfies the same evolution equation 
as the unpolarized PDF\cite{Aybat:2011ge}, its evolution is given by 
\begin{align}
f'^{\perp g}_{1T}(x,b;Q_f)=&\frac{M_p b}{2}T_{g,F}(x,x,Q_i)\exp\left\{-\int_{Q_i}^{Q_f}\frac{d\mu}{\mu}\left(A\ln\frac{Q^2}{\mu^2}+B\right)\right\}\left(\frac{Q_f^2}{Q_i^2}\right)^{-D(b^*;Q_i)} \nonumber\\ 
\times&\exp \left\{-b^2\left(g_1^\text{sivers}+\frac{g_2}{2}\ln\frac{Q_f}{Q_i}\right)\right\}
\end{align}
where $T_{q,F}(x,x,Q)$ is  the twist three  Qui-Sterman quark gluon correlation  function which is related  to the first $k_\perp$ 
moment of quark Sivers function\cite{Kang:2011mr} and can be expressed in terms of the unpolarized collinear PDFs~\cite{Kouvaris:2006zy,Echevarria:2014xaa}.
\begin{align}
T_{q,F}(x,x,Q)=\mathcal{N}_q(x)f_{q/P}(x,Q)
\end{align}
The expansion coefficients with the appropriate gluon anomalous dimensions at
NLL, $A^{(1)}, A^{(2)}, B^{(1)}$ and $D^{(1)},$  are known\cite{Echevarria:2014xaa}. 
Choosing the initial scale $Q_i=c/b$, the $D$ term vanishes at NLL. Taking Fourier transform of Eq. (9), one gets 
$f^{\perp g}_{1T}(x,k_\perp;Q_f)$ which is related to Sivers function through
\begin{equation}
\Delta^{N}f_{g/p^{\uparrow}}(x_{g},\bfk_{\perp g},Q)=-2\frac{k_{\perp g}}{M_p}f^{\perp g}_{1T}(x_g,k_{\perp g};Q)\cos\phi_{k_\perp}
\end{equation}

\section{Numerical Estimates of Asymmetry using  analytical and exact formalisms}
We will now present our estimates of SSA in photoproduction of $J/\psi$ for JLAB, HERMES, COMPASS and 
eRHIC energies. A detailed discussion of results can be found in Ref. \refcite{Godbole:2014tha}.
Figs. 1-5 show the y and $k_T$ distribution for  different experiments  with parameterizations TMD Exact-1, TMD Exact -2 and TMD a given in Table \ref{parameter:set}.  
TMD-e1 parameter set, extracted at $Q_0 = 1.0\text{ GeV}$, is for the exact solution of TMD evolution equations extracted in Ref. \refcite{Anselmino:2012aa}. TMD-a is the parameter set extracted using an analytical approximated form of evolution equations given in Ref. \refcite{Anselmino:2012aa}.
For estimates using NLL kernel we have used the most recent parameters by Echevarria {\it et al.}\cite{Echevarria:2014xaa}
obtained by performing a global fit of all experimental data on Sivers asymmetry in SIDIS from HERMES, COMPASS 
and JLAb. We call this set TMD-e2. This set was fitted at $Q_0 = \sqrt{2.4}\text{ GeV}$. 
Fig. 6 shows a comparison of asymmetries at all energies.

 \section{Summary}
We have compared estimates of SSA in electroproduction of $J/\psi$ using  TMD's evolved via DGLAP evolution and TMD evolution schemes. 
For the latter, we have chosen three different parameter  sets fitted using  an approximate analytical solution, an exact solution at LL and an exact solution at NLL. 
We find that the estimates given by TMD evolved PDF's and Sivers function are all comparable but substantially small as compared to estimates calculated using DGLAP evolved TMD's.
 
\begin{table}[t]
\begin{center}
  % \begin{tabular}{ |c|c|c| } 
\begin{tabular}{ | l | l | l | p{5cm} |}
 \hline
 TMD-e1 & TMD-a & TMD-e2 \\ 
 \hline 
$N_u=0.77, N_d=-1.00$&$N_u=0.75, N_d=-1.00$&$N_u=0.106,N_d=-0.163$\\
$a_u=0.68,a_d=1.11$&$a_u=0.82,a_d=1.36$&$a_u=1.051,a_d=1.552$\\
$b_u=b_d=3.1, $&$b_u=b_d=4.0, $&$b_u=b_d=4.857$, \\
$M_1^2=0.40\text{GeV}^2$&$M_1^2=0.34\text{GeV}^2$&$\langle k^2_{s\perp}\rangle = 0.282 \text{ GeV}^2$\\
$\langle k^2_{\perp}\rangle = 0.25\text{GeV}^2$&$\langle k^2_{\perp}\rangle = 0.25\text{GeV}^2$&$\langle k^2_{\perp}\rangle = 0.38\text{GeV}^2$\\
$b_{max} = 0.5 GeV^{-1}$&$b_{max} = 0.5 GeV^{-1}$&$b_{max} = 1.5 GeV^{-1}$\\
$g_2 = 0.68 \text{ GeV}^2$&$g_2 = 0.68 \text{ GeV}^2$&$g_2 = 0.16 \text{ GeV}^2$\\
 \hline
\end{tabular}
\caption{Parameter sets}
 \label{parameter:set}
\end{center}
\end{table}

\begin{figure}[h]
\begin{center}
\includegraphics[width=0.37\linewidth,angle=0]{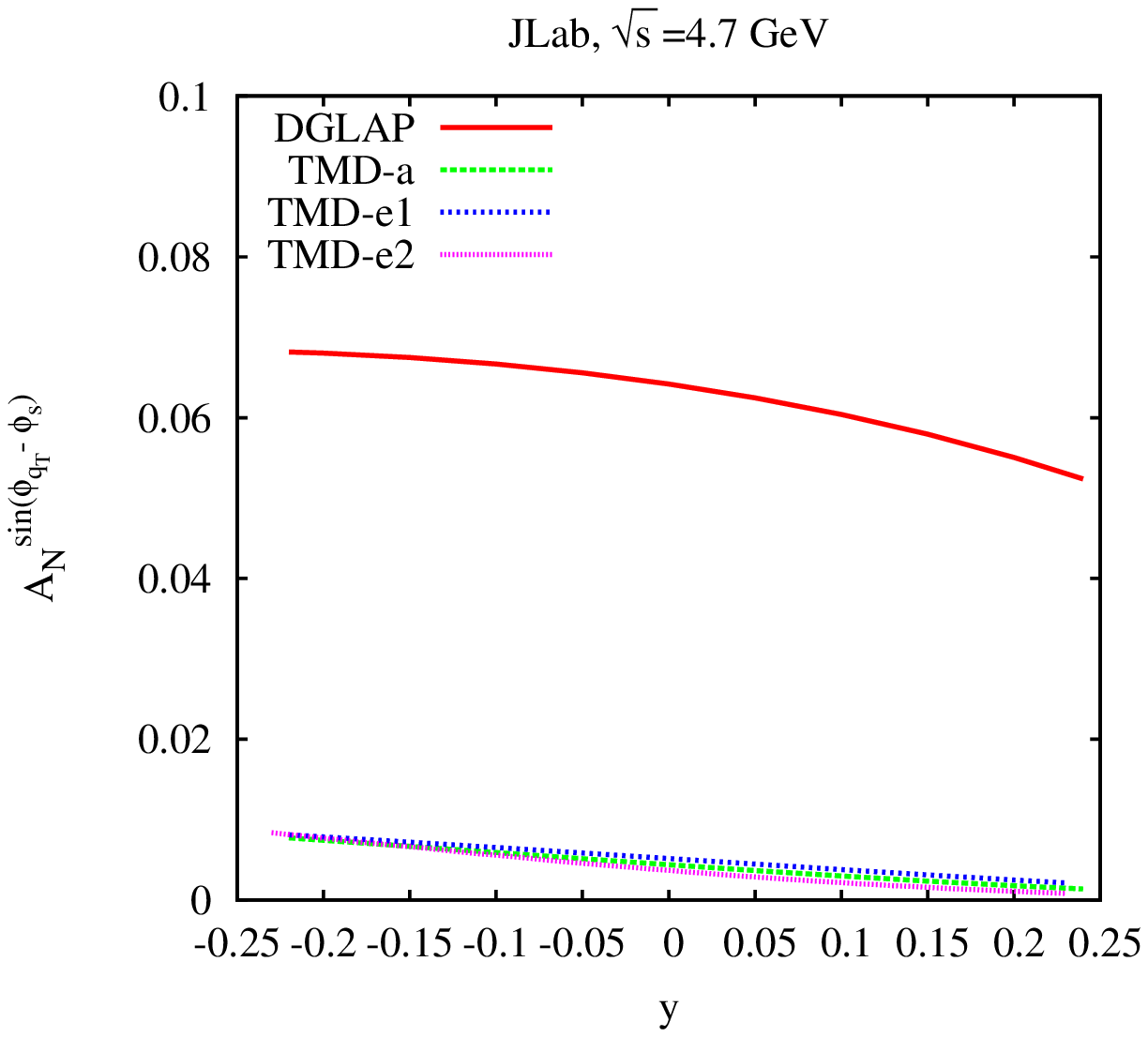}
\includegraphics[width=0.37\linewidth,angle=0]{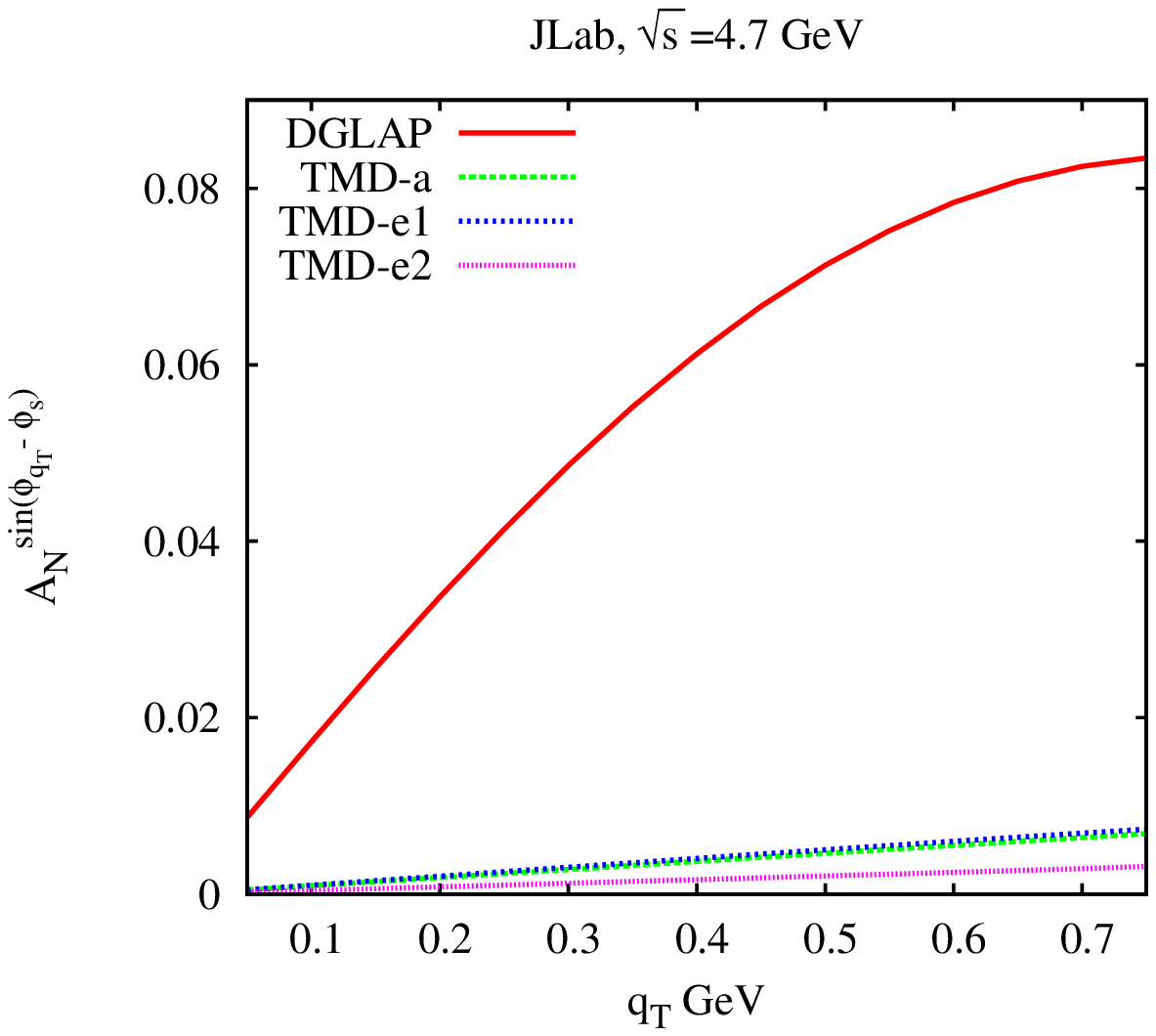}
\caption{The Sivers asymmetry $A_N^{\sin({\phi}_{q_T}-\phi_S)}$
 for $e+p^\uparrow \rightarrow  e+J/\psi +X $
at JLab energy ($\sqrt{s} = 4.7$ GeV), as a function of $y$ (left panel) and $q_T$ (right panel).
 The integration ranges are $(0 \leq q_T \leq 1)$~GeV and $(-0.25 \leq y \leq 0.25)$.}. 
\label{jlab_a}
\end{center}
\end{figure}
\begin{figure}[h]
\begin{center}
\includegraphics[width=0.37\linewidth,angle=0]{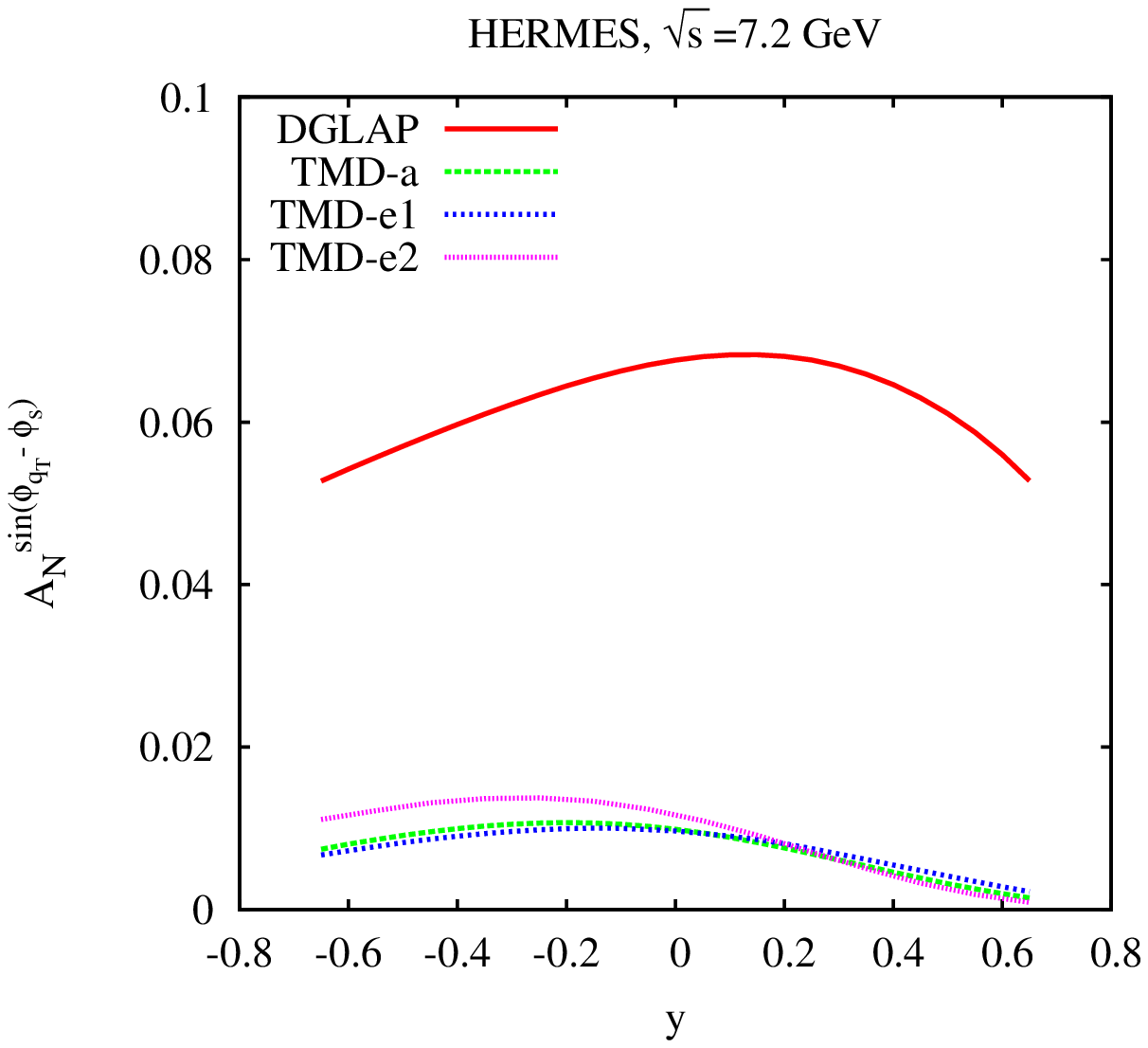}
\includegraphics[width=0.37\linewidth,angle=0]{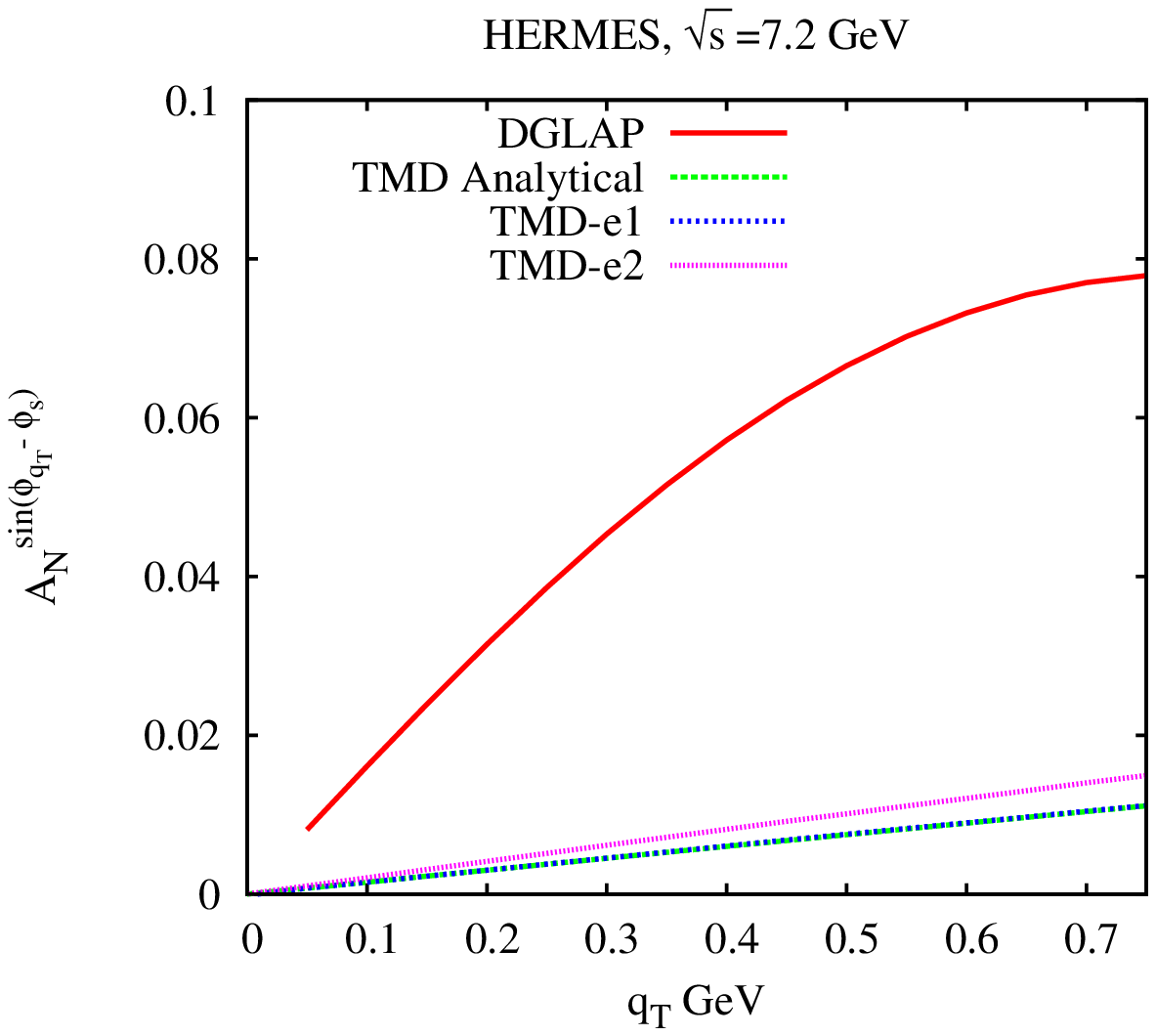}
\caption{ HERMES energy ($\sqrt{s} = 7.2$ GeV), Asymmetry as a function of $y$ (left panel) and $q_T$ (right panel).
 The integration ranges are $(0 \leq q_T \leq 1)$~GeV and $(-0.6 \leq y \leq 0.6)$.}
\label{hermes_a}
\end{center}
\end{figure}
\begin{figure}[h]
\begin{center}
\includegraphics[width=0.37\linewidth,angle=0]{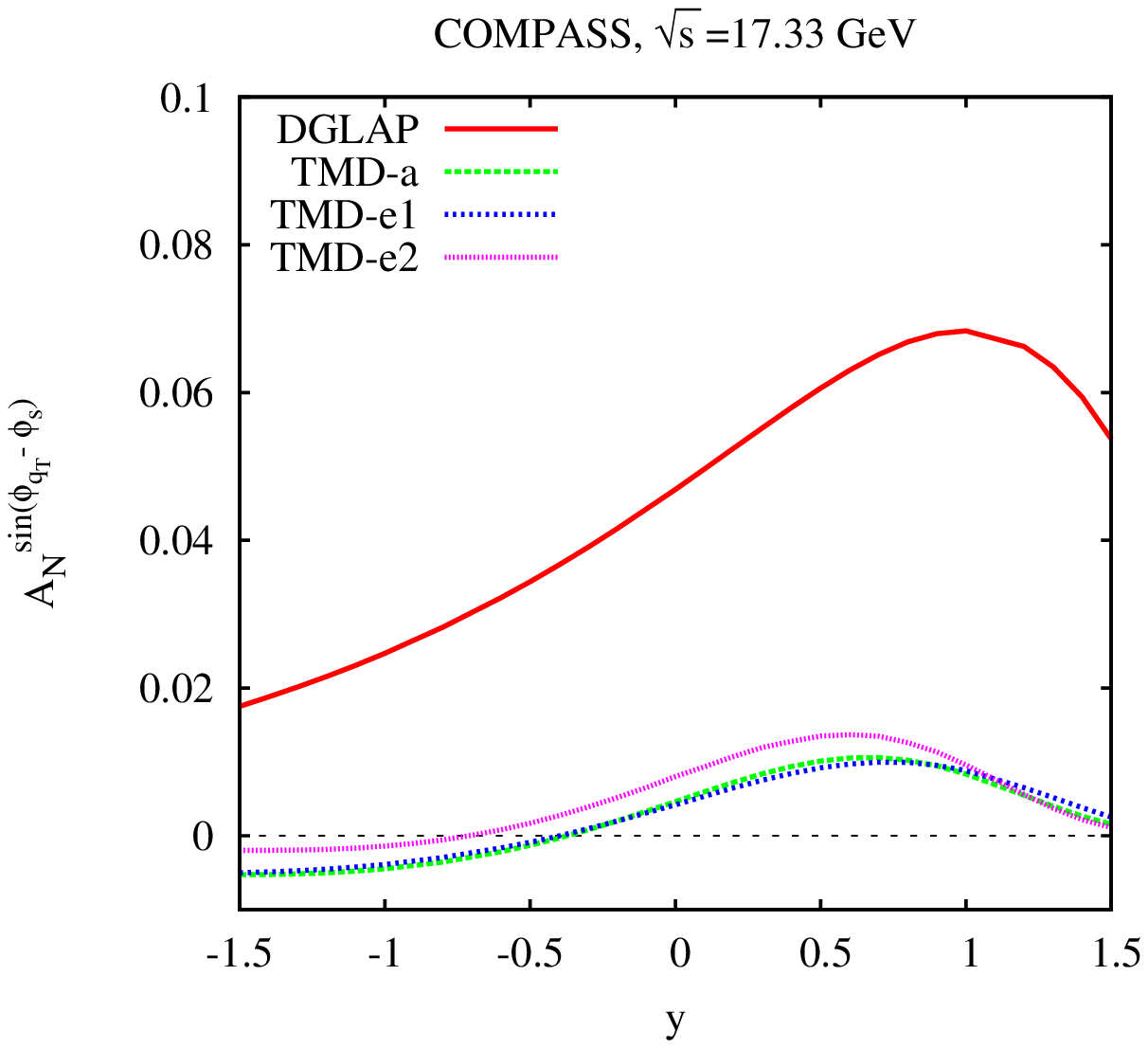}
\includegraphics[width=0.37\linewidth,angle=0]{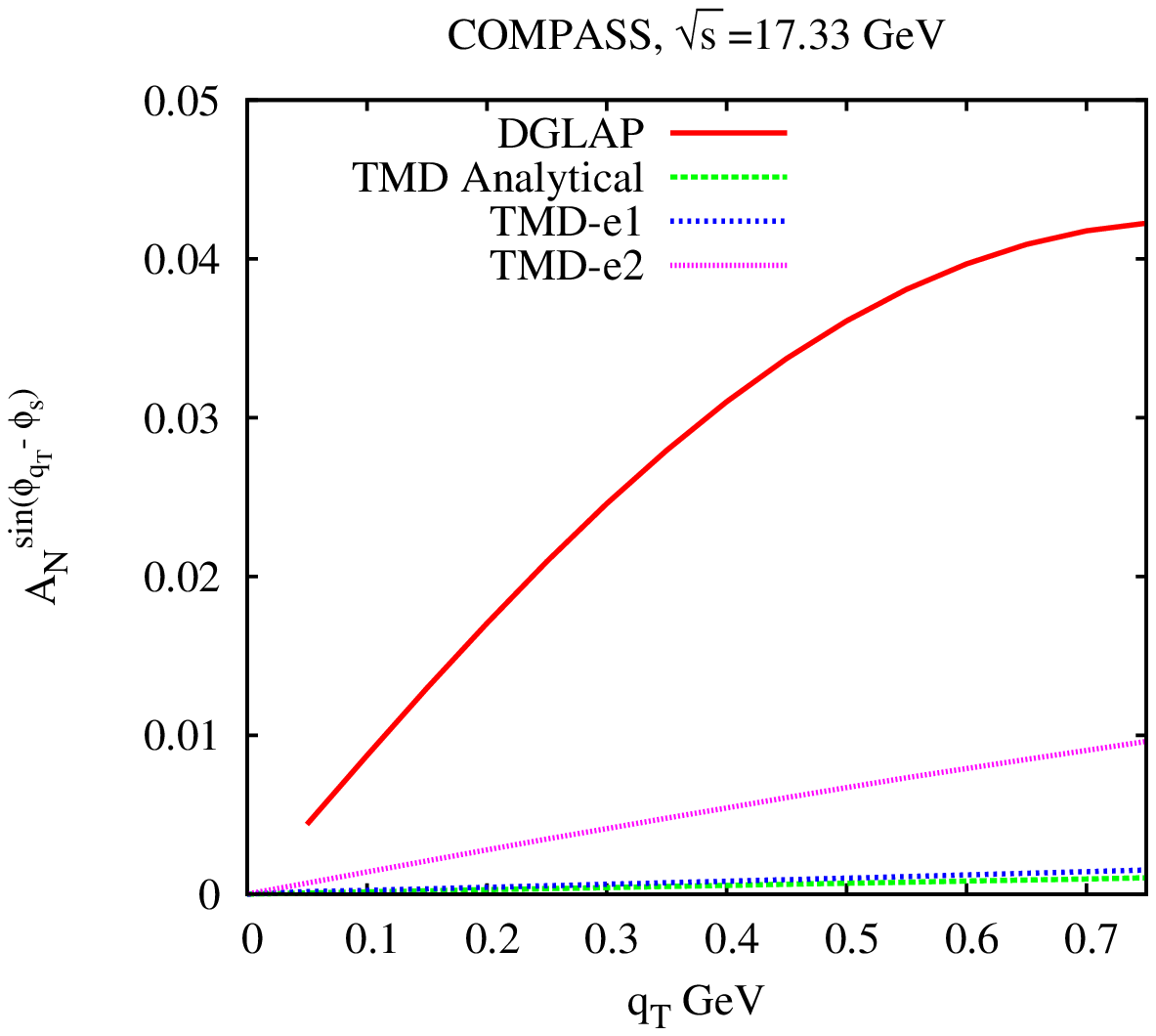}
\caption{COMPASS energy ($\sqrt{s} = 17.33$ GeV),Asymmetry as a function of $y$ (left panel) and $q_T$ (right panel).
 The integration ranges are $(0 \leq q_T \leq 1)$ GeV and $(-1.5 \leq y \leq 1.5)$.}
\label{compass_a}
\end{center}
\end{figure}
\begin{figure}[h]
\begin{center}
\includegraphics[width=0.37\linewidth,angle=0]{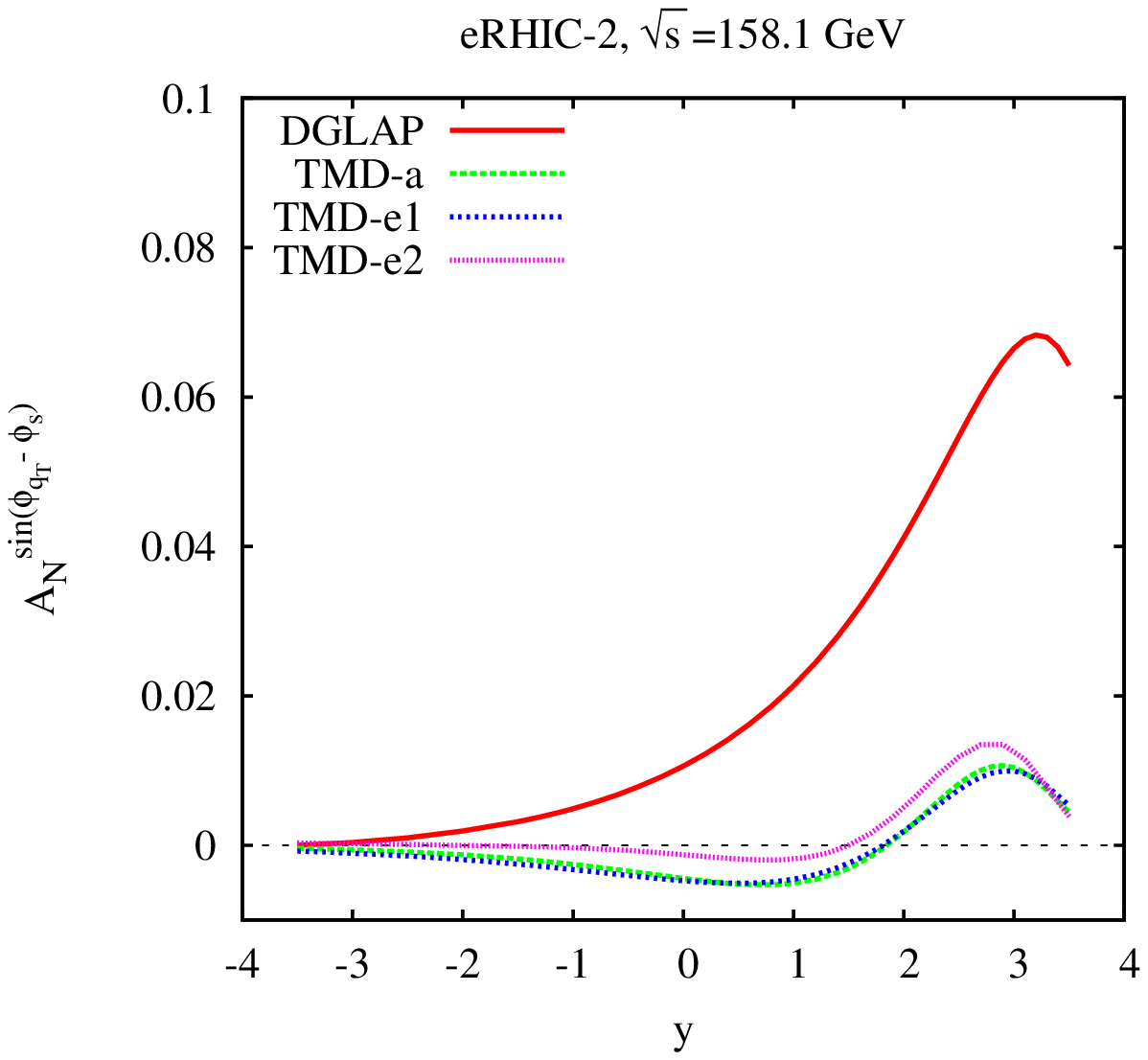}
\includegraphics[width=0.37\linewidth,angle=0]{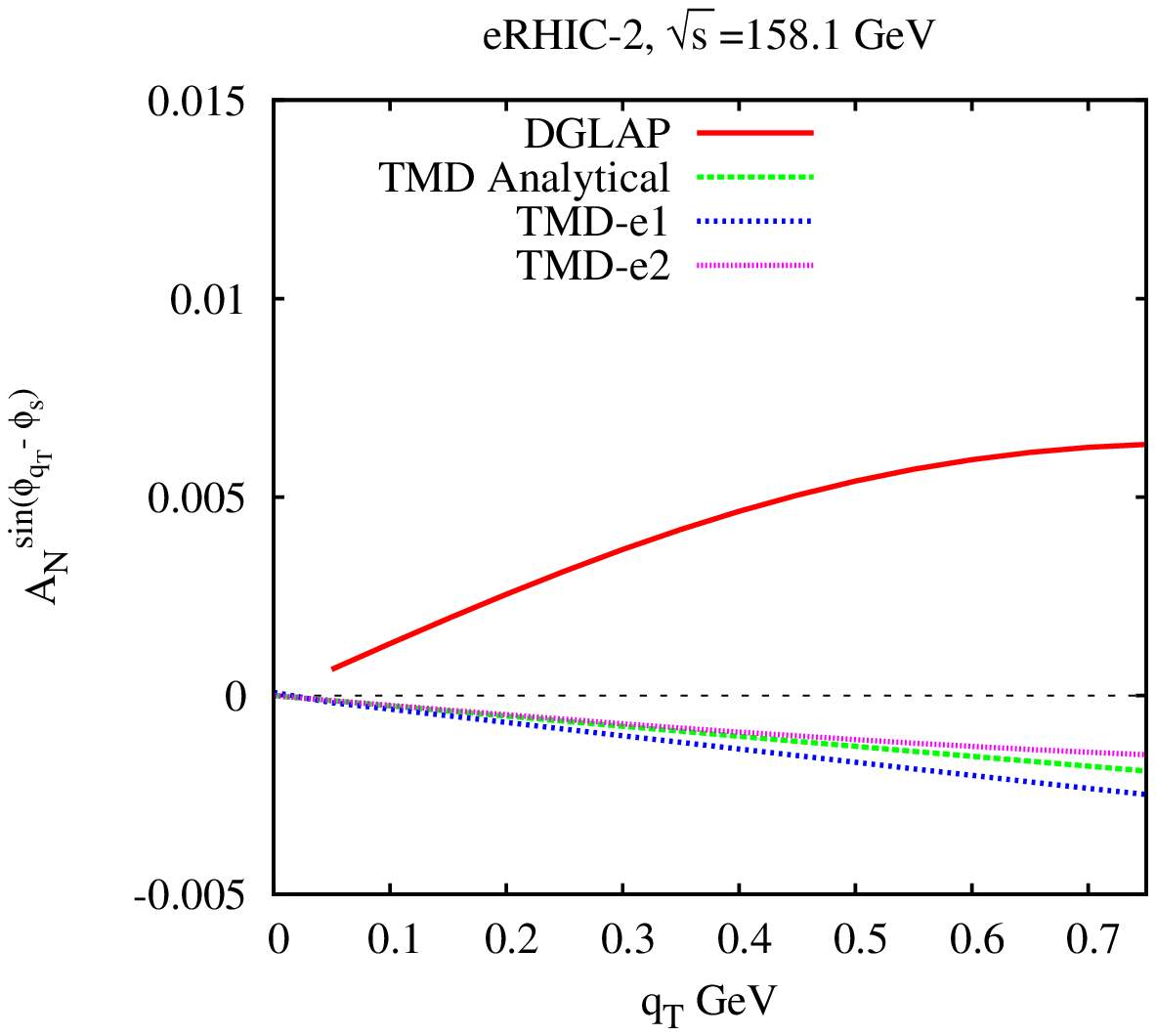}
\caption{eRHIC energy ($\sqrt{s} = 158.1$ GeV),Asymmetry as a function of $y$ (left panel) and $q_T$ (right panel).
. The integration ranges are $(0 \leq q_T \leq 1)$ GeV and $(-3.7 \leq y \leq 3.7)$. }
\label{erhic2_a}
\end{center}
\end{figure}
\begin{figure}[h]
\begin{center}
\includegraphics[width=0.37\linewidth,angle=0]{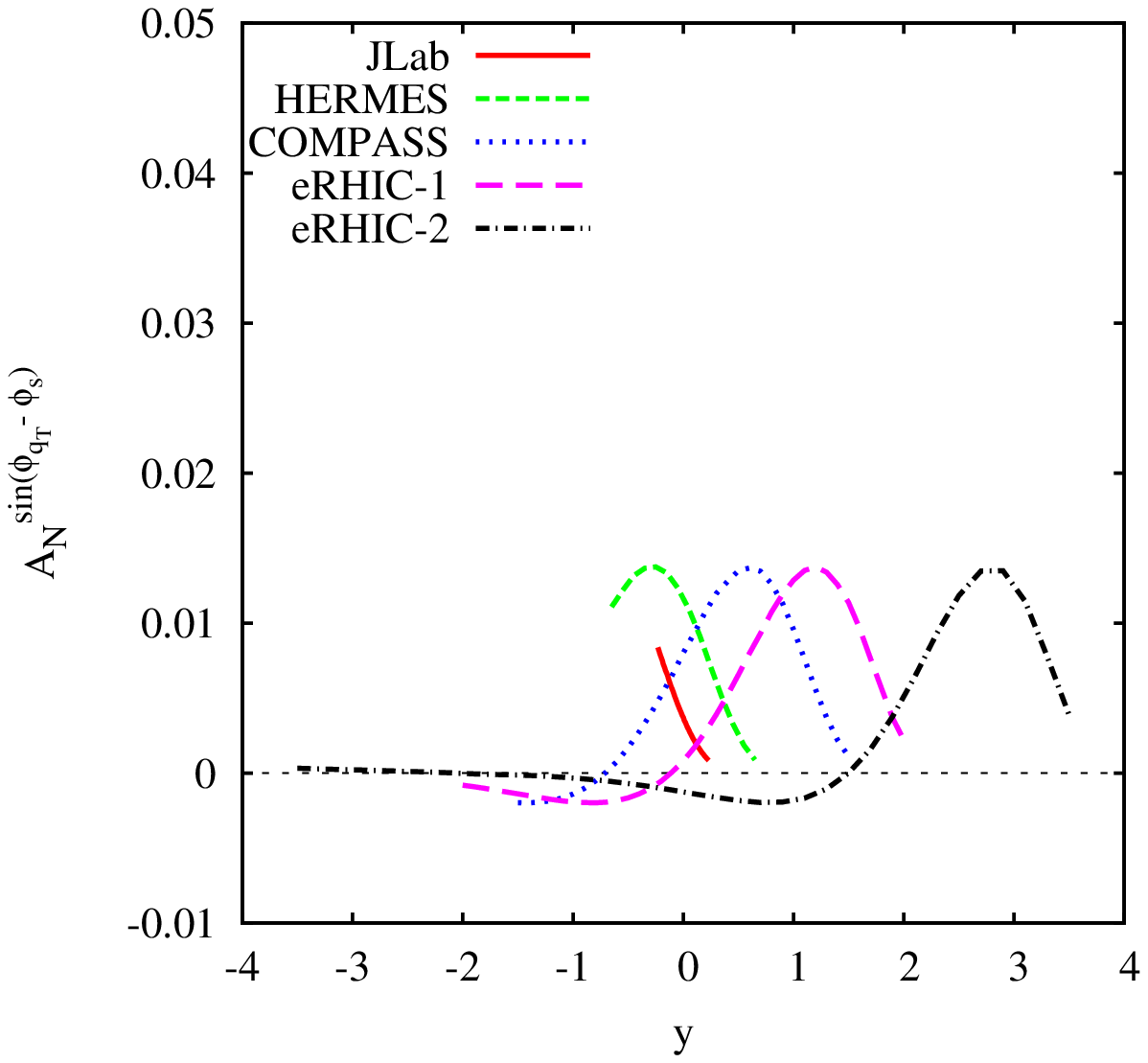}
\includegraphics[width=0.37\linewidth,angle=0]{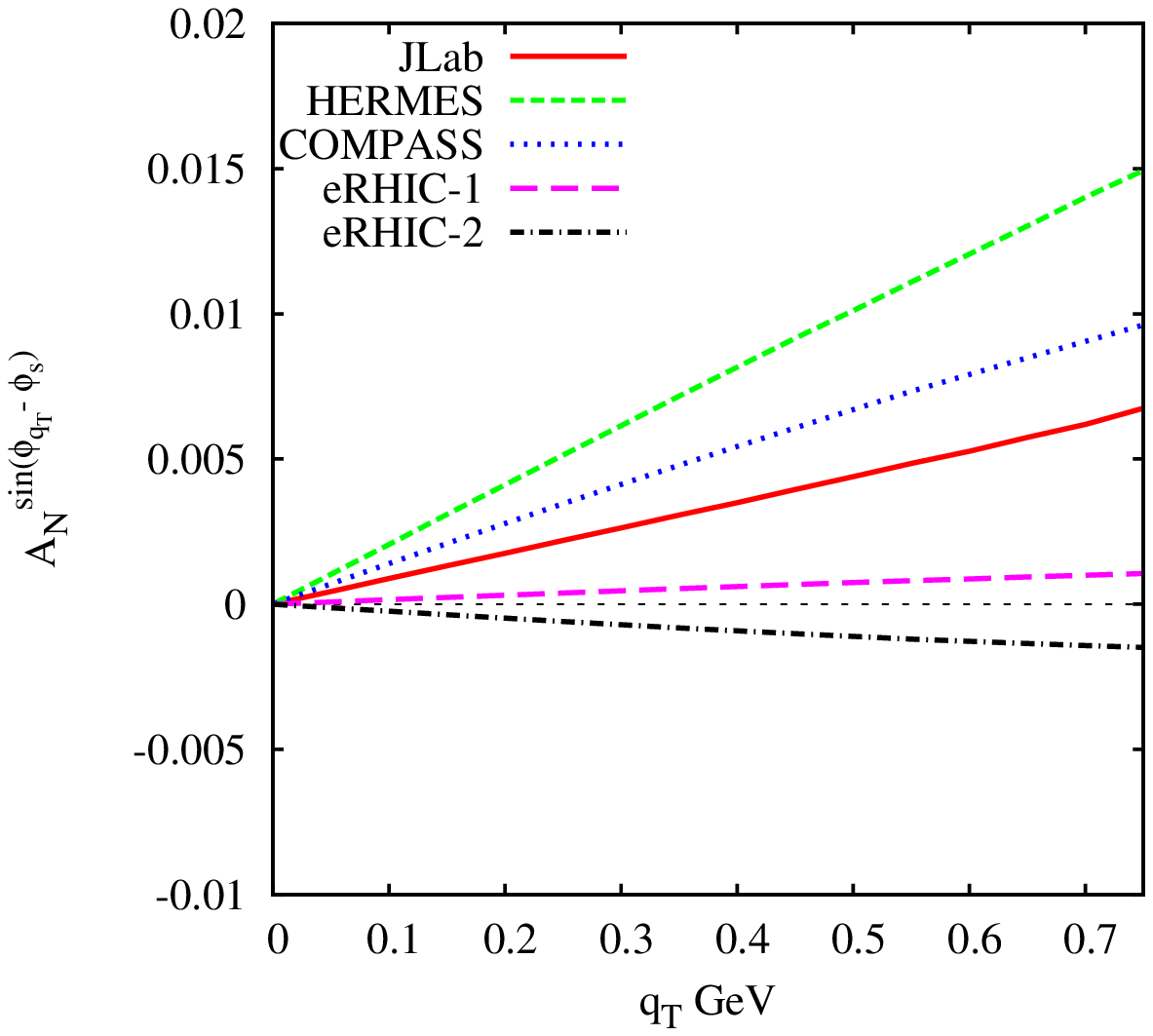}
\caption{Left panel: Plot of the Sivers asymmetries in the $y$ distribution obtained in all c.o.m energies using the TMD-e2 fit . This plot shows the drift of the asymmetry peak towards higher values of rapidity $y$. Right panel: Plot of the Sivers Asymmetries in the $q_T$ distribution\label{compare_y}}
\end{center}
\end{figure}

\clearpage
\section*{Acknowledgements}
AM would like to thank the organizers of QCD2014 for their warm hospitality and 
University of Mumbai, University Grants Commission, India and Indian National Science Academy for financial support. This work was done 
under Grant No 2010/37P/47/BRNS of Department of Atomic Energy, India.

\end{document}